\documentclass[letterpaper]{JHEP3}

\usepackage{epsfig}
\usepackage{amsmath}
\usepackage{xspace}

\def\be{\begin{equation}}
\def\ee{\end{equation}}
\def\beq{\begin{eqnarray}}
\def\eeq{\end{eqnarray}}
\def\a{\alpha}
\def\b{\beta}

\def\l{\lambda}

\def\({\left (}
\def\){\right )}
\def\[{\left [}
\def\[{\right ]}

\def\ra{\rightarrow}

\title{Towards a Novel no-hair Theorem for Black Holes}
\author{Thomas Hertog \\
Theory Division, CERN, CH-1211 Geneva 23, Switzerland \\
{\it and}\\
APC, 11 Place Marcelin Berthelot, 75005 Paris, 
France \thanks{UMR 7164 (CNRS, Universit\'e Paris 7, CEA, Observatoire de Paris)}\\
E-mail: \email{Thomas.Hertog@cern.ch}}

\date{July 2006}

\abstract{%

We provide strong numerical evidence for a new no-scalar-hair theorem for black holes in general 
relativity, which rules out spherical scalar hair of static four dimensional black holes if the 
scalar field theory, when coupled to gravity, satisfies the Positive Energy Theorem. This sheds light 
on the no-scalar-hair conjecture for Calabi-Yau compactifications of string theory, where the effective 
potential typically has negative regions but where supersymmetry ensures the total energy is always 
positive. In theories where the scalar tends to a negative local maximum of the potential at infinity,  
we find the no-scalar-hair theorem holds provided the asymptotic conditions are invariant under the full 
anti-de Sitter symmetry group.

}

\preprint{CERN-PH-TH/2006-163}

\begin{document}

%
\section{Introduction}
%

Inspired by the uniqueness theorems for static \cite{Israel67} and stationary \cite{Carter71} asymptotically flat vacuum black holes in Einstein-Maxwell theory, Wheeler famously conjectured  \cite{Ruffini71} that `black holes have no hair' in more general (sensible) matter theories, in four dimensions. Following Wheeler's conjecture a number of rigorous no-hair theorems - which however are manifestly limited in scope - were established \cite{Chase70,Teitelboim72,Bekenstein72}. On the other hand, Wheeler's hypothesis was proven wrong in e.g. Einstein-Yang-Mills \cite{Bizon90} and Einstein-Skyrme \cite{Bizon92} theory, in various combinations with dilaton \cite{Lavrelashvili93} or Higgs \cite{Greene93} fields. Some, but not all, of these hairy black holes are unstable. For gravity coupled to scalar fields, however, possibly in combination with Abelian gauge fields, a precise formulation of the no-hair conjecture has not yet been given. In this arena there still remains interesting ground to explore between several rigorous (but limited) no-hair theorems, and a number of explicit hairy black hole solutions which appear to disprove Wheeler's conjecture in its most general form.

The first no-scalar-hair theorems applied to the massless scalar \cite{Chase70}, and to the neutral scalar field with a monotonically increasing self-interacting potential \cite{Bekenstein72}. They show there are no regular asymptotically flat spherical black hole solutions with scalar hair for minimally coupled scalar fields with convex potentials. These theorems were later generalized to the case of minimally coupled scalar fields with arbitrary positive potentials \cite{Heusler92}, and also to scalar multiplets \cite{Bekenstein95} and to nonminimally coupled scalars\footnote{The BBM black hole \cite{Bocharova70} and its magnetic monopole extension \cite{Virbhadra94} provides, strictly speaking, a counterexample to the no-hair theorems for conformally coupled scalars. This hardly compromises the spirit of the no-hair conjectures, however, because the hair amounts to a discrete parameter \cite{Xanthopoulos91}, and the black hole is known to be linearly unstable \cite{Bronnikov78}.} \cite{Mayo96}. Using similar techniques, it was shown more recently \cite{Sudarsky02} there are no hairy asymptotically anti-de Sitter (AdS) black holes where the scalar field asymptotically tends to the global (negative) minimum of the potential.

Recent developments in string theory, however, indicate there is little or no justification to restrict attention to positive scalar potentials, or to asymptotic conditions defined with respect to the global minimum of the scalar potential. Potentials with a local negative maximum which fall off exponentially are familiar indeed in supersymmetric AdS compactifications. Similarly, a large class of supersymmetric string theory compactifications of the form $M_4 \times K$, where $M_4 $ is four dimensional Minkowski space and $K$ is a Ricci flat, compact manifold admitting a covariantly constant spinor, give rise to effective four dimensional potentials with negative regions \cite{Hertog03}. These include compactifications on all simply connected Calabi-Yau and $G_2$ manifolds. The underlying mathematical reason for this is that all simply connected compact manifolds of dimension five, six or seven admit Riemannian metrics with positive scalar curvature. From an effective four dimensional standpoint, positive scalar curvature on $K$ acts as negative energy density\footnote{The effective potentials are generally unbounded from below, since one can rescale the metric on $K$ and make the scalar curvature arbitrarily large.}. 

On the other hand, the Positive Energy Theorems (PET) ensure that the total ADM energy remains positive in supersymmetric string theory compactifications \cite{Witten81,Gibbons83}. This can be understood as a generalization of the well-known phenomenon that gravity can stabilize a false vacuum \cite{Coleman80}. 
For Calabi-Yau compactifications, this means that the negative regions of the effective four dimensional potentials must be separated by a positive barrier from the local minimum at zero, which corresponds to a metric on $K$ on the moduli space. In other words the metrics with positive scalar curvature on 
$K$ lie a finite distance away from the moduli space of Ricci flat metrics. The PET ensures that in all (nonsingular) asymptotically flat solutions, the positive energy density of the barrier `screens' and cancels out any negative energy density of the central regions of solutions, rendering the total energy positive. 

It is therefore natural to ask whether the no-scalar-hair conjecture holds in supersymmetric string theory compactifications. To gain some insight in this problem we examine here the no-hair conjecture for gravity coupled to scalar field theories with potentials that are qualitatively similar to those that arise in string theory compactifications. 
It is widely believed that spherical asymptotically flat black holes with scalar hair generally exist when the scalar potential has negative regions\footnote{See e.g. \cite{Gubser05} for explicit examples.}. This is due, in part, to the proliferation in recent years of neutral \cite{Torii01,Hertog04} (and charged \cite{Martinez06}) asymptotically AdS black hole solutions with scalar hair, where the scalar asymptotically tends to a local negative extremum of the potential. When this is a maximum (which we assume satisfies the Breitenlohner-Freedman (BF) bound \cite{Breitenlohner82})) the scalar generically decays slower than usual. Nevertheless, it has been verified that some of the asymptotic black hole solutions preserve the full AdS symmetry, and that the conserved charges associated with the asymptotic symmetries are well-defined and finite \cite{Hertog04,Henneaux04}. 

However, a careful analysis of the scalar field theories that admit hairy black hole solutions reveals that none of them satisfies the PET. One finds that either the potential barrier around the local extremum is too low to compensate for all the negative energy density \cite{Gubser05}, or that the (AdS) black holes obey unusual boundary conditions for which there are solutions\footnote{These need not be black holes.} with negative total energy \cite{Torii01,Hertog04}. Based on this and on extensive numerical evidence that we present below, we conjecture that {\it there are no static asymptotically flat and asymptotically AdS black holes with spherical scalar hair if the scalar field theory, when coupled to gravity, satisfies the Positive Energy Theorem.} 

In other words we argue that, rather than the presence of a cosmological constant or the fact that the scalar potential has negative regions, the nonperturbative stability of the ground state is the only relevant feature of the theory that determines whether or not the no-scalar-hair conjecture holds.
This applies to all scalar potentials (bounded and unbounded) with a local minimum at zero, and to all potentials with a negative local extremum. When the latter is a local maximum at or slightly above the BF bound, however, one must require the asymptotic conditions are invariant under the full AdS symmetry group. 
Indeed, we show hairy black holes do exist in certain stable `designer gravity' theories \cite{Hertog05}, where one considers AdS gravity coupled to tachyonic scalars with boundary conditions that break some of the AdS symmetries.

A brief outline of this paper is as follows. In section 2, we consider gravity minimally coupled to a scalar field and we review what are the necessary and sufficient conditions on the scalar potential for the theory to satisfy the PET. This enables us in section 3 to construct a large class of `critical potentials' which are on the verge of violating the PET. In section 4 we provide strong numerical evidence that scalar theories which are critical in this sense, also separate the set of theories where the no-scalar-hair theorem (and the PET) hold from those where it does not. These numerical calculations are performed for a wide range of theories, with asymptotically flat and with asymptotically AdS boundary conditions, which should be representative for all scalar field theories with asymptotic conditions specified with respect to a local (or global) minimum of $V$. In section 5 we extend these findings to scalar field theories where $\phi$ reaches a negative maximum of $V$ at infinity (for all choices of AdS-invariant boundary conditions). We also study the no-hair conjecture in designer gravity theories, where the asymptotic conditions break the conformal symmetry. We find a branch of hairy black holes in a consistent truncation of ${\cal N}=8$ $D=4$ gauged supergravity with boundary conditions for which the AdS/CFT duality \cite{Aharony00} indicates the gravity theory should satisfy the PET. We conclude in section 6 with some comments on possible generalization of our results, and the significance of no-hair theorems in general. The details of the numerical calculations that support the extension of the no-scalar-hair conjecture to potentials with negative regions are given in Appendix A.

%
\section{Positive Mass}
%

We consider gravity in $d \geq 4$ spacetime dimensions minimally coupled to a scalar 
field with potential $V(\phi)$. So the action is
\be \label{act}
S=\int d^d x \sqrt{-g} \left [{1\over 2} R - {1\over 2}(\nabla\phi)^2 -V(\phi)\right ]
\ee
where we have set $8\pi G=1$. We require the potential can be written as
\be \label{superpot}
V(\phi) = (d-2) P'^2 - (d-1)P^2
\ee
for some function $P(\phi)$. Scalar potentials of this form arise in the context of $N=1$ supergravity 
coupled to $N=1$ matter, in which case $P(\phi)$ is the superpotential. We are interested in configurations 
where $\phi$ asymptotically approaches a local extremum of $P$ at $\phi=\phi_0$. An extremum 
of $P$ is always an extremum of $V$, and $V (\phi_0) \leq 0$. Hence configurations where $\phi \ra \phi_0$ at infinity correspond to asymptotically flat or asymptotically anti-de Sitter (AdS) solutions.  

At an extremum of $P$ one has
\be\label{ddsuperpot}
V'' = 2P'' \left[ (d-2)P'' -(d-1)P\right]
\ee
so a local extremum of $P$ corresponds to a minimum of $V$ except when $0<(d-2)P'' <(d-1) P $, or when  
$(d-1) P < (d-2)P'' <0$. This is a quadratic equation for $P''$, which has a real solution if and only if
\be
V'' \geq - {(d-1)^2 P^2 \over 2(d-2)} = {(d-1)V (\phi_0) \over 2(d-2)} = - {(d-1)^2 \over 4l^2}
\ee
where $l^2$ is the AdS radius. Hence we recover the BF bound $m^2_{BF}= - {(d-1)^2 \over 4l^2}$ on the scalar mass, which is required for AdS solutions to be perturbatively stable. 

By generalizing Witten's spinorial proof of the Positive Energy Theorem (PET) for asymptotically flat 
spacetimes \cite{Witten81}, Townsend has shown \cite{Townsend84} (see also \cite{Gibbons83}) that for 
the case of a single scalar field a potential $V(\phi)$ admits a PET {\it if and only if} $V$ can be written in terms of a 'superpotential' $P(\phi)$, and $\phi$ asymptotically decays (sufficiently fast\footnote{Requiring $\phi$ approaches an extremum of $P$ at infinity uniquely specifies the asymptotic behavior of the fields, except when $\phi_0$ corresponds to a negative local maximum of $V$. In this case $\phi$ generically decays as a combination of two modes. The PET 
\cite{Townsend84} holds, in general, only for scalar boundary conditions that select the mode with the faster fall-off 
\cite{Hertog06}. This is what we mean here by decaying `sufficiently fast'.  The connection between the PET and the validity of the no-hair theorem for different choices of scalar boundary conditions is analysed in section 5.}) to an extremum of $P$. 

The argument in \cite{Townsend84} obviously concerns the positivity properties of the energy as defined by the spinor charge\footnote{Townsend's theorem also establishes the positivity of the conserved energy associated with the Hamiltonian generator 
${\cal H}_{\partial_{t}}$, because this equals the spinor charge when $\phi \rightarrow \phi_0$ sufficiently fast.} 
\be \label{spinorcharge}
{\mathcal Q}_{\partial_{t}} = \oint *{\bf B}
\ee 
where the integrand is the dual of the Nester 2-form, with components
\be\label{2form}
{B}_{ab} = 
\frac{1}{2}(\overline \psi \gamma^{[c} \gamma^d \gamma^{e]} \widehat \nabla_e \psi 
+ {\rm h.c.})\epsilon_{abcd} \, 
\ee
$\psi$ is taken to be an asymptotically supercovariantly constant spinor field and
\be\label{covder}
\widehat \nabla_a \psi 
= \left[ \nabla_a - {1 \over \sqrt{2(d-2)}} \gamma_a P(\phi) \right]\psi
\ee
with $P(\phi)$ given by (\ref{superpot}). This definition of the covariant derivative enabled \cite{Townsend84} to 
express the spinor charge (\ref{spinorcharge}) as a manifestly non-negative quantity, provided $\psi$ satisfies the spatial Dirac equation $\gamma^i \widehat D_i \, \psi = 0$. In the context of $N=1$ supergravity $P$ is the superpotential, but the argument of \cite{Townsend84} applies to any gravity scalar theory, irrespective of whether it is a sector of a supergravity theory.

%
\section{Critical Potentials}
%

Townsend's result \cite{Townsend84} implies that {\it all} potentials that are on the verge of violating the PET must correspond to functions $V$ that are on the boundary of when one can solve (\ref{superpot}) for $P$. This provides a clear criterion for a potential 
to be critical in this sense, which can be expressed as a local analytic condition on $P$ as follows.

\begin{figure}[htb]
\begin{picture}(0,0)
\put(257,198){V}
\put(325,105){$\phi$}
\put(47,105){$\phi_0$}
\end{picture}
\centering{\psfig{file=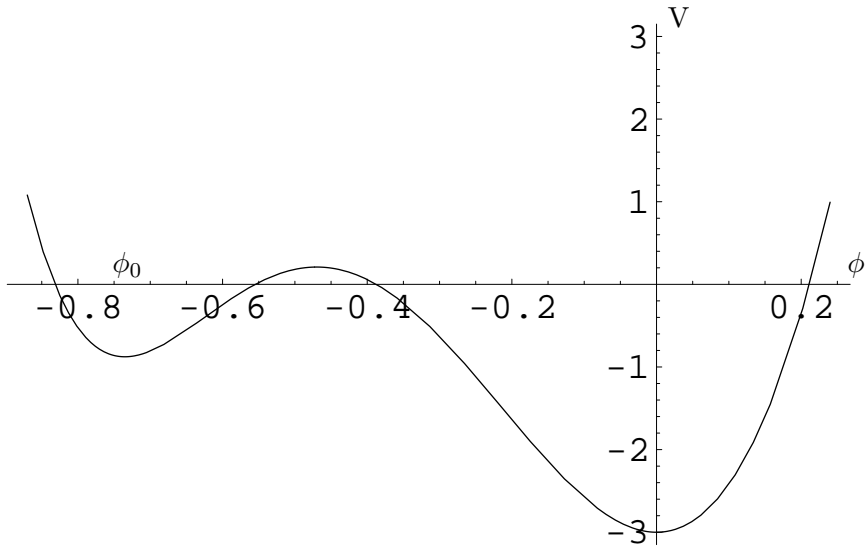,width=4.5in}}
\caption{A potential $V(\phi)$ that is on the verge of violating the Positive Energy Theorem for solutions that asymptotically approach the local AdS minimum at $\phi_0$.}
\label{1}
\end{figure}

First consider potentials with a local extremum at $\phi_0$ where $V \leq 0$ and $V'' \geq m^2_{BF}$, and with a global
minimum at $\phi =0$. An example is shown in Figure 1. To construct $P$ we try to solve
\be\label{Weq}
P'(\phi) =\frac{1}{\sqrt{d-2}}\sqrt{V+(d-1) P^2} 
\ee
starting with $P = \sqrt{-V/(d-1)}$ at $\phi_0$.

A solution to (\ref{Weq}) exists unless the quantity inside the square root becomes negative. As we integrate 
out from $\phi_0$, $P$ is increasing and the square root remains real because the scalar satisfies the BF bound. 
Hence if the global minimum at $\phi=0$ is not very much lower than the local extremum at $\phi_0$, a global solution for 
$P$ will exist and $P'(0)>0$. This is expected, since the PET holds for potentials of this form. 
If the global minimum is too deep, however, the quantity under the square root will become negative before 
the global minimum is reached, and a real solution will not exist. Clearly the critical potential corresponds 
to one where $V+(d-1) P^2$ just vanishes as the global minimum is reached. In other words, the condition\footnote{We have verified that this condition agrees with the criterion proposed in \cite{Hertog03b}, which states that the volume contribution to the total mass of a certain class of spherical time-symmetric initial data vanishes for critical potentials.} for a potential $V$ to be on the verge of violating the PET is simply $P'(0)=0$. This applies to all
critical potentials that are bounded from below, since every potential for which the theory (\ref{act}) satisfies the PET is of the form (\ref{superpot}) for some $P$.

\begin{figure}[htb]
\begin{picture}(0,0)
\put(35,200){B}
\put(200,140){I}
\put(200,90){II}
\put(253,115){III}
\put(282,152){$B_f$}
\put(310,138){$B_c$}
\put(253,95){IV}
\put(253,50){V}
\put(325,22){A}
\end{picture}
\centering{\psfig{file=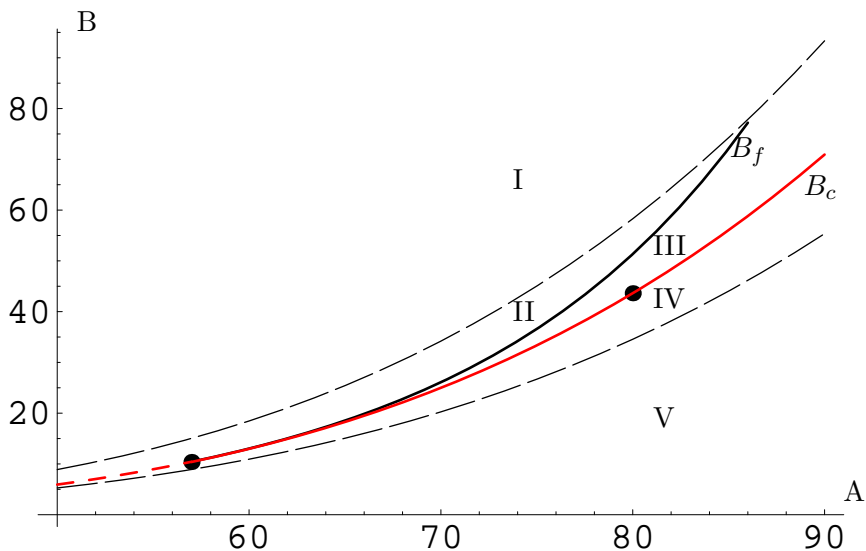,width=4.5in}}
\caption{The function $B_c(A)$ represents a set of critical potentials of the form (3.2), with a local AdS minimum at $\phi_0 <0$ and a global minimum at $\phi=0$. The function $B_f(A)$ corresponds to potentials with a local minimum at $\phi_0$ where $V(\phi_0)=0$. The combination $(A,B)=(57.6,10.87)$ where $B_c=B_f$ yields a critical potential with a local minimum at zero.}
\label{2}
\end{figure}

This yields a simple prescription for constructing critical potentials. As an illustration, consider potentials of the following form in four dimensions,
\be\label{V1}
V(\phi) = -3 +50 \phi^2 +A \phi^3 + B\phi^6
\ee
where $A$ and $B$ are free parameters. For combinations $(A,B)$ in region III and IV in Figure 2, these are qualitatively similar to the potential shown in Figure 1. That is, $V$ has a local minimum at $\phi_0 <0$ where $V \leq 0$, and a global minimum at $\phi=0$. For combinations $(A,B)$ in region II one has $V>0$ at $\phi_0$, and in region I there is no local extremum at $\phi_0$ at all. Finally, for combinations of parameters in region V one has 
$V(\phi_0) <V(0)$.

We have numerically solved (\ref{Weq}) for a range of parameters, tuning the combination $(A,B)$ such that $P'(0)=0$. This yields a one-parameter family of critical potentials\footnote{A similar class of critical potentials exists in region V for configurations where asymptotically $\phi \ra 0$.}, given by the function $B_c(A)$ that separates region III from region IV in Figure 2. Potentials in region III do not satisfy the PET for solutions where $\phi \rightarrow \phi_0$ at infinity, whereas potentials in region IV do admit the PET for these asymptotic conditions. The critical potential shown above in Figure 1 corresponds to the marked point at $A=80$ on $B_c$. Potentials where $V(\phi_0)=0$ are represented by the function $B_f(A)$ in Figure 2. This function joins\footnote{The dashed continuation of $B_c$ toward smaller values of $A$ represents theories where $V (\phi_0)=0$, which are critical only in the sense that they separate potentials with an unstable de Sitter false vacuum at $\phi_0$ (in region II) from potentials with a stable AdS false vacuum (in region IV).} $B_c$ at $(A,B)=(57.6,10.8718)$, which thus yields a potential that is on the verge of violating the PET with asymptotically flat boundary conditons $\phi \rightarrow \phi_0$.

\begin{figure}
\begin{picture}(0,0)
\put(288,15){$\Lambda$}
\put(285,175){$h$}
\end{picture}
\centering{\psfig{file=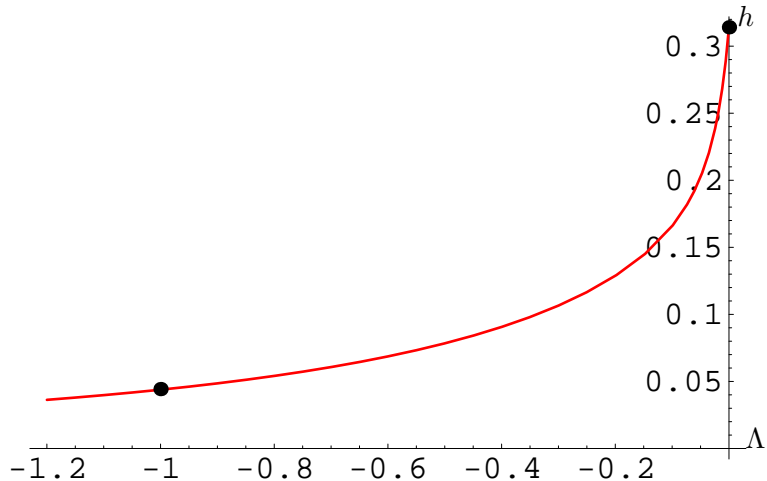,width=4in}}
\caption{The function $h_{c} (\Lambda)$ that specifies a set of critical potentials of the form (3.3), which are unbounded from below.}
\label{3}
\end{figure}

We have not been able to find a simple local condition on $P$ which characterizes critical potentials that are unbounded from below\footnote{This would necessarily involve the shape of $V$ at large $\phi$ \cite{Hertog03}.}, but one can still solve (\ref{Weq}) to generate numerically a large set of (approximately) critical potentials of this type. As before, Townsend's result implies that 
{\it all} unbounded potentials $V$ which are on the verge of violating the PET can be constructed this way.

Consider e.g. the following class of potentials,
\be \label{V2}
V(\phi) = \Lambda +\phi^2 + C\phi^3
\ee
where $\Lambda \leq 0$ is a cosmological constant, and $C$ controls the height $h= 4/27 C^2$ of the barrier around the local minimum at $\phi=0$. One can solve (\ref{Weq}) for $P$, starting with $P'=0$ at $\phi=0$ and tuning $(\Lambda,C)$ so that a global solution $P(\phi)$ just exists. This yields a class of critical potentials, where the barrier around the local minimum at $\phi=0$ is just high enough to ensure all configurations where $\phi \rightarrow 0$ at infinity have positive total mass. These critical theories can be specified by a function $h_c(\Lambda)$, which we plot in Figure 3. The critical potential for $\Lambda=0$ is shown in Figure 4.

\begin{figure}
\begin{picture}(0,0)
\put(227,177){V}
\put(290,102){$\phi$}
\end{picture}
\centering{\psfig{file=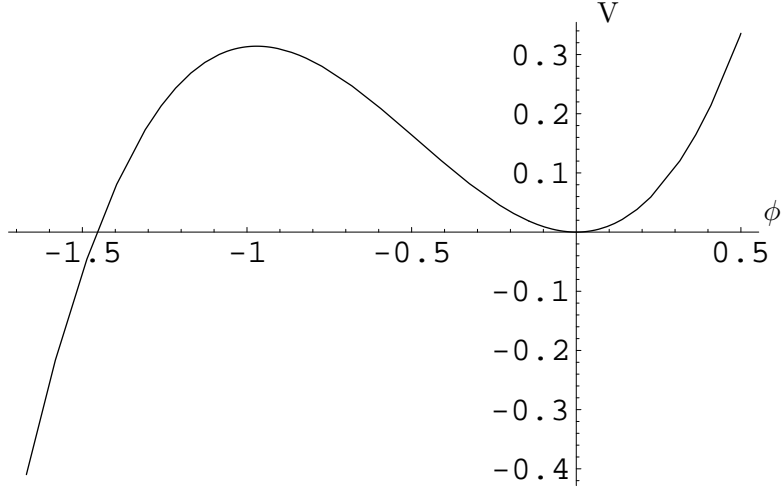,width=4in}}
\caption{A potential $V(\phi)$ that is on the verge of violating the the Positive Energy 
Theorem for solutions that asymptotically approach the local minimum at $\phi =0$.}
\label{4}
\end{figure}

%
\section{Black Holes with Scalar Hair}
%

We now turn to the no-scalar-hair conjecture that we have proposed, which states that in theories of gravity coupled to a single scalar with potential $V$, the PET holds if (and only if) the theory admits no regular static black hole solutions with spherical scalar hair. Writing the metric for static spherical solutions in $d$ spacetime dimensions as
\be 
ds_d^2=-h(r)e^{-2\chi(r)}dt^2+h^{-1}(r)dr^2+r^2d\Omega_{d-2}^2
\ee
the Einstein equations read
\be \label{fieldeq1}
h\phi_{,rr}+\left(\frac{(d-2)h}{r}+\frac{r}{(d-2)}\phi_{,r}^2h+h_{,r} 
\right)\phi_{,r} =V_{,\phi}
\ee
\be \label{fieldeq2}
(d-3)(1-h)-rh_{,r}-\frac{r^2}{(d-2)}\phi_{,r}^2h={ 2 \over (d-2)}r^2V(\phi)
\ee
\be \label{fieldeq3}
\chi_{,r} = - \frac{r \phi_{,r}^2}{(d-2)}
\ee
At the event horizon $R_e$ one has $h=0$. Regularity at $R_e$ requires 
\be \label{horcon}
\phi'(R_{e}) = {V_{,\phi} \over h_{,r}} = {(d-2)R_e V_{,\phi_{e}} \over (d-2)(d-3)-2R_e^2 V(\phi_e)}
\ee
where $\phi_{e}\equiv \phi(R_e)$. Rescaling $t$ shifts $\chi$ by a constant, so its value at the event horizon is arbitrary. 
Asymptotically we require $\phi $ tends (sufficiently fast) to an extremum of $V$ at $\phi_0$, with $V(\phi_0) \leq 0$.
So the metric function $h$ can be written as
\be \label{asmetric5d}
h(r)={ r^2 \over l^2}+1-\frac{m(r)}{r^{d-3}} 
\ee
where $l^2$ is the AdS radius when $V(\phi_0)<0$. Substituting this form of $h$ in the field equations allows us to integrate (\ref{fieldeq2}), which yields the following expression for the mass\footnote{The mass (\ref{spinorcharge}) recieves an extra finite contribution from the scalar field if this saturates the BF bound \cite{Hertog03c}.} (\ref{spinorcharge}),
\beq \label{bhmass}
M & = & {(d-2) \over 2} {\mathrm Vol}(S^{d-2}) \lim_{r \rightarrow \infty} m(r) \nonumber\\
& = &  e_{\ }^{-F_{\infty}} \left[ M_s +{\mathrm Vol}(S^{d-2}) \int_{R_e}^{\infty} e^{F(\tilde r)}
\left( V(\phi)-\Lambda + {1 \over 2} \left(1+ {\tilde r^2 \over \ell^2}\right) \phi_{,\tilde r}^2 \right) \tilde r^{d-2} d\tilde r \right] \nonumber\\
\eeq
where ${\mathrm Vol}(S^{d-2})$ is the volume of a unit $(d-2)$-sphere, and  
$F (r)={1\over d-2} \int_{R_e}^{r}  d \hat  r \ \hat r \phi_{,\hat r}^2$ with  $F_{\infty} \equiv F(r=\infty)$. The Schwarschild-(AdS) black hole mass $M_s$ is given by
\beq
M_s= {(d-2) \over 2} Vol(S^{d-2}) \left({1 \over l^2} R_e^{d-1}  + R_e^{d-3} \right)
\eeq
We emphasize that regular spherical black hole solutions with scalar hair would be specified by a single charge. The existence of spherical hairy black holes with positive mass would therefore provide a genuine counterexample to the no-scalar-hair theorem that we propose, because there would be a vacuum Schwarschild (or Schwarschild-AdS) black hole with 
$\phi(r)=\phi_0$ everywhere and with the same mass.

One can verify whether black holes with scalar hair exist in a given theory by numerically integrating (\ref{fieldeq1})-(\ref{fieldeq3}) outward from the horizon for a range of $R_e$ and $\phi_e$, and see if there exists a combination $(R_e,\phi_e)$ for which $\phi \rightarrow \phi_0$ at infinity. We have done this analysis for theories of the form (\ref{V1}) and (\ref{V2}), where $\phi_0$ corresponds to a local minimum of $V$.
In particular, we have integrated the field equations for several one-parameter families of potentials, labeled by $\lambda$, that are represented by functions $B_{\lambda}(A)$ (in Fig. 2) and $h_{\lambda}(\Lambda)$ (in Fig. 3) that intersect the curves $B_c$ and $h_c$ of critical potentials at $\lambda =0$. In other words, the scalar field theories represented by $B_{\lambda}$ and $h_{\lambda}$ admit the PET for, say $\lambda  \geq 0$, whereas for $\lambda <0$ the PET does not hold. In Appendix A we show that for all functions $B_{\lambda}(A)$ and $h_{\lambda}(\Lambda)$ of this form, $\lambda <0$ if and only if there exists (precisely one) value $\phi_e \neq \phi_0$ (for each horizon size $R_e$) for which the scalar profile obeys the prescribed asymptotic conditions. Regular black hole solutions with scalar hair cease to exist when $\lambda \rightarrow -0$, either because $\phi_e $ reaches the global minimum in this limit or because $\phi_e \rightarrow \infty$ (for all $R_e$). We refer the reader to Appendix A for the details of these numerical calculations.

Since potentials of this form should be representative for all (single) scalar field theories with asymptotic conditions specified with respect to a local (or global) minimum of $V$, this strongly indicates that the no-scalar-hair theorem holds in theories that satisfy the PET, and it shows that black holes with scalar hair exist if $V$ does not admit the PET. We emphasize that our analysis includes bounded and unbounded potentials, with asymptotically flat as well as with asymptotically AdS boundary conditions. In the next section we argue these findings extend to all scalar field theories where $\phi$ reaches a negative maximum of $V$ at infinity, provided the asymptotic conditions preserve the full AdS symmetry group.

%
\section{Designer Gravity}
%

When $\phi_0$ corresponds to a negative local maximum of $V$, requiring $\phi \ra \phi_0$ at infinity does not uniquely specify its asymptotic profile. This is already evident from the linearized wave equation $\nabla^2 \delta \phi - m^2 \delta \phi=0$ for tachyonic scalars in an AdS background. Solutions with harmonic time dependence $e^{-i\omega t}$ decay asymptotically as\footnote{For fields that saturate the BF bound, $\l_{+}=\l_{-} \equiv \l$ and $\delta \phi = {\alpha \over r^{\lambda}}\ln r  + {\beta \over r^{\lambda}}$.}
\be\label{genfall}
\delta \phi \sim  {\alpha \over r^{\lambda_{-}}}  + {\beta \over r^{\lambda_{+}}}
\ee
where $\a$ and $\b$ are functions of $t$ and the angles and
\be\label{fallofftest}
\lambda_\pm = {d-1 \over 2}  \pm  {1 \over 2} \sqrt{(d-1)^2 + 4l^2 m^2}
\ee
For scalar masses in the following range
\be \label{range}
m^2_{BF} \leq m^2 < m^2_{BF}+{ 1 \over l^2}<0
\ee
both modes are normalizable, and hence represent a priori physically acceptable fluctuations. 

To have a well-defined theory one must specify boundary conditions at spacelike infinity. In general this amounts to choosing a functional relation between $\a$ and $\b$ in (\ref{genfall}). The standard choice of boundary condition, for which Townsend's Positive Energy Theorem applies, corresponds to taking $\a=0$. When one takes in account the self-interaction of the scalar field, as well as its backreaction on the geometry, one finds this is consistent with the usual set of asymptotic conditions on the metric components that is left invariant under $SO(d-1,2)$ \cite{Henneaux85}.

To get an idea whether the no-scalar-hair conjecture that we have proposed extends to theories of this form we consider the following class of potentials in four dimensions,
\be
V(\phi) = -3 -\phi^2 - D \phi^4
\ee
where $D$ is a free parameter. The scalar generically decays as $\phi \sim \alpha/r + \beta/r^2$ at infinity. Solving (\ref{Weq}) for a range of $D$ reveals that the PET holds for $\a=0$ boundary conditions provided $D <3/4$. In Appendix A we show that this parameter regime coincides precisely with the regime where there exists no finite (nonzero) value of the scalar field at the horizon for which the asymptotic $1/r$ mode is switched off completely. Instead, for all $\phi_e$ the scalar asymptotically behaves as a combination of the two modes. By contrast, when $D>3/4$ there is always one value $\phi_e$ for which asymptotically $\phi \sim 1/r^2$.
Hence it appears that at least with standard boundary conditions on tachyonic scalars in AdS, which select the mode with the faster fall-off, the PET forbids spherical scalar hair of black holes\footnote{The analysis in \cite{Hertog06} of spherical soliton solutions in a wide range of models where $\phi_0$ corresponds to a maximum of $V$ provides further support for the validity of the no-hair conjecture proposed here.}.

What is, however, the status of the no-scalar-hair theorem for different scalar boundary conditions, defined by $\a \neq 0$ and $\b = \b(\a)$? The backreaction of the $\a$-branch of the scalar field (as well as its self-interaction) modify the asymptotic behavior of the gravitational fields, but it has recently been shown that the Hamiltonian generators of the asymptotic symmetries remain well-defined and finite when $\a \neq 0$ \cite{Hertog04,Henneaux04,Hertog05}. The generators acquire, however, an explicit contribution from the scalar field. In particular, the conserved mass 
${\cal H} [\partial_t] $ of spherical solutions is given by
\be\label{mass}
M = {\rm Vol}(S^{d-2}) \left[{(d-2) \over 2} M_0 + \lambda_{-} \a\b +(\lambda_{+} - \lambda_{-}) W \right]
\ee
where $M_0$ is the coefficient of the $1/r^{d+1}$ term in the asymptotic expansion of $g_{rr}$, and where we have defined the function
\be \label{bc}
W(\a) = \int_0^\a \b(\tilde \a) d \tilde \a 
\ee
which defines the choice of boundary conditions.
These so-called `designer gravity' boundary conditions (\ref{bc}) generally break the AdS symmetry to $\Re \times SO(d-1)$, since the asymptotic scalar profile changes under the action of $\xi^r$. The full AdS symmetry group is preserved, however, for asymptotic conditions defined by
\be \label{adsW}
W(\a)=k \a^{d-1/\lambda_{-}}
\ee
where $k$ is an arbitrary constant without variation\footnote{The function $W$ that specifies AdS-invariant boundary conditions
is more complicated at certain discrete values of the scalar mass \cite{Henneaux04}.}.

The dynamical properties of the theory - including the possible formation of hairy black holes - depend significantly on the choice of $W$. Townsend's Positive Energy Theorem \cite{Townsend84} need not hold with designer gravity boundary conditions \cite{Hertog06}, but the AdS/CFT correspondence indicates there is a lower bound on the conserved energy in those designer (super)gravity theories that, for $W=0$, admit a dual description in terms of a supersymmetric conformal field theory (CFT).
In the context of the AdS/CFT correspondence, adopting more general scalar boundary conditions defined by a function $W \neq 0$ corresponds to adding a potential term $\int W ({\cal O})$ to the dual CFT action, where ${\cal O}$ is the field theory operator that is dual to the bulk scalar \cite{Witten02, Berkooz02}. 
Certain deformations $W$ give rise to field theories with additional, possibly metastable vacua. The AdS/CFT correspondence asserts that the expectation values $\langle {\cal O} \rangle$ in different field theory vacua can be obtained from regular static solitons in the bulk. The precise correspondence between solitons and field theory vacua is given by the following function \cite{Hertog05},
\be \label{effpot}
{\cal V}(\alpha) = -\int_{0}^{\a} \b_{s} (\tilde \a) d\tilde \a + W(\alpha)
\ee
where $\b_{s}(\a)$ is obtained from the asymptotic scalar profile of soliton solutions for different values of 
$\phi$ at the origin. It has been shown \cite{Hertog05} that for any $W$ the location of the extrema of ${\cal V}$ yield the vacuum expectation values $ \langle {\cal O} \rangle = \a$, and that the value of ${\cal V}$ at each extremum yields the energy of the corresponding soliton. Hence ${\cal V}(\a)$ can be interpreted as an effective potential for $\langle {\cal O} \rangle$.

This led \cite{Hertog05} to conjecture that (a) there is a lower bound on the gravitational energy in those designer gravity theories where ${\cal V}(\a)$ is bounded from below, and that (b) the solutions locally minimizing the energy are given by the spherically symmetric, static soliton configurations found in \cite{Hertog05}. It would follow in particular that the true vacuum of the theory is given by the lowest energy spherical soliton. For ${\cal V}(\a) \geq 0$ everywhere one recovers the usual positivity properties of the energy, with perfect AdS as the ground state of the theory.

Here we are concerned with the connection between the nonperturbative stability of designer gravity theories and the validity of the no-scalar-hair theorem. Hairy black holes have been found in designer gravity \cite{Torii01,Hertog04,Martinez06}, for AdS-invariant boundary conditions (\ref{adsW}) as well as for other choices of $W(\a)$. In all known examples, however, the boundary conditions render the usual AdS vacuum unstable. This suggests that the PET excludes spherical scalar hair in designer gravity, in line with our results for standard 
$\a=0$ boundary conditions. We now show, however, that the situation in designer gravity is slightly more subtle.

Consider gravity minimally coupled to a single scalar with potential
\be\label{potsugra}
V(\phi) = -2-\cosh(\sqrt{2}\phi)
\ee
This is a consistent truncation of ${\cal N}=8$ supergravity in four dimensions, which arises as the low energy limit of M theory compactified on $S^7$. The potential has a maximum at $\phi=0$ corresponding to an $AdS_4$ solution with unit radius. It is unbounded from below, but small fluctuations have $m^2 =-2$, which is slightly above the BF bound in $d=4$.  
Hence in all asymptotically AdS solutions $\phi$ decays at large radius as
\be\label{genfall2}
\phi = {\alpha \over r}  + {\beta \over r^2}
\ee
where $r$ is an asymptotic area coordinate, and $\b(\a)$ in designer gravity. 
Writing the metric as $g_{\mu \nu}=\bar g_{\mu \nu} +h_{\mu \nu}$ where $\bar g_{\mu \nu}$ is the metric of perfect AdS spacetime, the corresponding asymptotic behavior of  the metric components is given by
\be \label{4-grr}
h_{rr}=  -{(1+\alpha^2/2) \over r^4} + {\cal O} (1/r^5), \quad h_{rm} = {\cal O}(1/r^2), \quad  h_{mn}={\cal O}(1/r)
\ee
If we adopt boundary conditions defined by the function
\be \label{bccurve}
\b_{bc}(\a) = - {1 \over 5} \a^2 +{1 \over 30} \a^3
\ee
the conserved mass (\ref{mass}) of spherical solutions is given by
\be \label{bhmass2}
M = 4\pi \(M_0 - { 15 \over 4} \a^3 + {1 \over 6} \a^4\)
\ee

\begin{figure}[htb]
\begin{picture}(0,0)
\put(2,70){$\beta$}
\put(206,138){$\alpha$}
\put(259,139){$M/4\pi$}
\put(434,16){$R_e$}
\end{picture}
\mbox{
\epsfxsize=7.1cm\epsfysize=5cm \epsffile{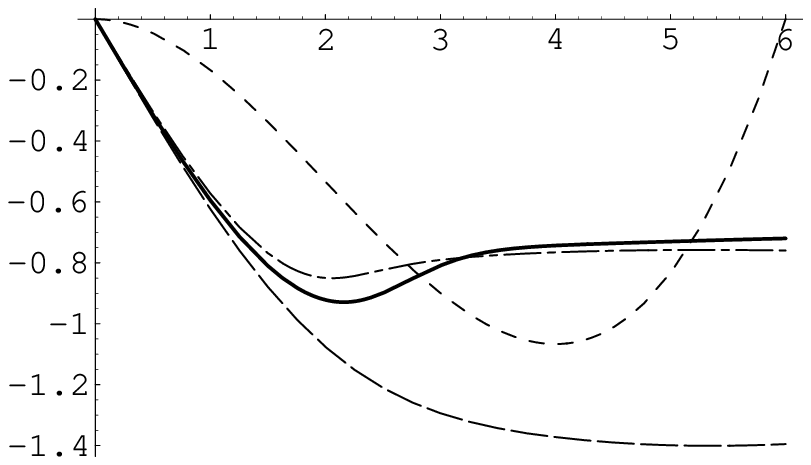} \qquad 
\epsfxsize=6.9cm \epsfysize=5cm \epsffile{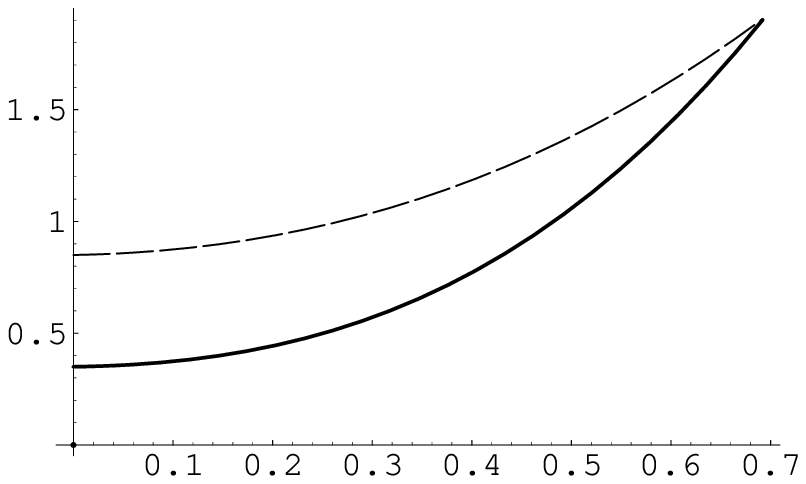}}
\vspace{.5cm}
\caption{The left panel shows the functions $\b(\a)$ obtained from the solitons and from hairy black holes of two different sizes. The full line shows the soliton curve $\b_{s}(\a)$, the dot-dashed line shows the $\b_{R_e}(\a)$ curve for $R_e=.2$ black holes and the dashed line is the $R_e=1$ curve. One sees the boundary condition function $\b_{bc}(\a)=-{1 \over 5} \a^2 +{1 \over 30} \a^3$, given by the dotted line, intersects the curves $\b_{R_e}(\a)$ twice for small $R_e$. The right panel shows the mass of the hairy black holes that obey these boundary conditions. The full line gives the masses of the second (perturbatively stable) branch of solutions, which are associated with the second intersection point of the curves $\b_{R_e}(\a)$ with $\b_{bc}(\a)$, and hence have more hair.}
\label{5}
\end{figure}

To find hairy black hole solutions we numerically integrate the field equations (\ref{fieldeq1})-(\ref{fieldeq3}) for static spherical
solutions outward from the horizon. The scalar asymptotically behaves as (\ref{genfall2}), so we obtain a point in the $(\a,\b)$ plane for each combination $(R_e,\phi_e)$. Repeating for all $\phi_e$ gives a curve $\b_{R_e}(\a)$. In Figure 5 (a) we show this curve for hairy black holes of two different sizes. As one increases $R_e$, the curve decreases faster and reaches larger (negative) values of $\b$. We also show the curve obtained in a similar way for regular solitons, which were discussed in 
\cite{Hertog05}. Finally, the dotted line corresponds to the boundary condition curve (\ref{bccurve}).

Given a choice of boundary conditions $\b(\a)$, the allowed black hole solutions are simply given by the points where the black hole curves intersect the boundary condition curve: $\b_{R_e}(\a) = \b(\a)$. 
It follows immediately that for designer gravity boundary conditions (\ref{bccurve}) there are two hairy black holes of a given horizon size if $R_e$ is sufficiently small. On the other hand, it is clear there are no large hairy black holes with these boundary conditions. The mass (\ref{bhmass2}) of both branches of solutions is plotted in Figure 5 (b). One sees all hairy black holes have positive mass, and in fact they are always more massive than a Schwarschild-AdS black hole of the same size. The hairy black holes provide a genuine example of black hole non-uniqueness, since for a given mass $M$ (below a critical value) there are three distinct black hole solutions.

From the soliton curve $\b_{s}(\a)$ one can compute the `effective potential' (\ref{effpot}) for the vacuum expectation values 
$\langle {\cal O} \rangle$, for any $W$. The result for boundary conditions (\ref{bccurve}) is plotted in Figure 6. 
One sees that ${\cal V} \geq 0$ everywhere. The AdS/CFT correspondence suggests, therefore, that the bulk theory (\ref{potsugra}) with boundary conditions (\ref{bc}) satisfies the PET, and hence that empty AdS remains the true ground state\footnote{See \cite{Hertog05c} for a stability analysis of this theory (with more stringent conditions on $W$) using purely gravitational arguments.}. 
The black hole solutions given here are, to our knowledge, the first examples of regular spherical black holes with scalar hair in a theory of gravity coupled to a single scalar field for which the PET should hold. We emphasize again, however, that the asymptotic solutions are not invariant under the full AdS symmetry group.

\begin{figure}[htb]
\begin{picture}(0,0)
\put(56,177){${\cal V}$}
\put(290,16){$\alpha$}
\end{picture}
\centering{\psfig{file=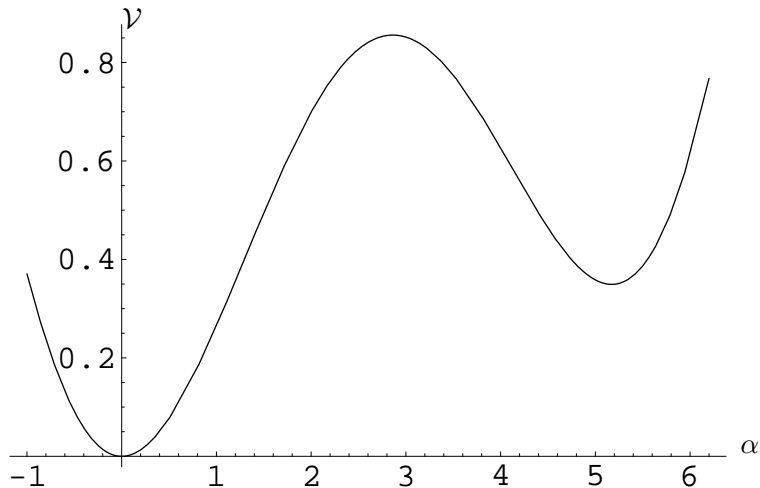,width=4in}}
\caption{The effective potential ${\cal V}(\a)$ for the vacuum expectation values $\langle {\cal O} \rangle$ in the dual field theory with deformation $W= - {1 \over 15} \a^3 + {1 \over 120} \a^4$.}
\label{6}
\end{figure}

Figure 6 shows that ${\cal V}$ has three extrema; a global minimum at $\a=0$ and two extrema at $\a \neq 0$ where 
${\cal V} >0$. Hence the dual field theory has three different vacua, and one can consider excitations about each. The usual Schwarschild-AdS black holes correspond to typical excitations of mass $M$ about the $\a=0$ vacuum. On the other hand, the hairy black holes are interpreted in the dual theory as thermal excitations about the metastable vacua with $ \langle {\cal O} \rangle \neq 0$ \cite{Hertog05}. In particular, the top branch of hairy black holes in Figure 5 (b) corresponds to excitations about the local maximum of ${\cal V}$, whereas the bottom branch corresponds to excitations about the local (metastable) minimum\footnote{This interpretation is supported by the fact that the top branch has an unstable spherically symmetric scalar perturbation, whereas the bottom branch does not \cite{Hertog05,Hertog04b}.}. The dual field theory description of hairy black holes, therefore, nicely resolves the black hole non-uniqueness in the bulk\footnote{The dual description also suggests one can view hairy black hole solutions of stable designer gravity theories somewhat as excited atoms in nonperturbative gravity. A thermodynamic analysis along the lines of \cite{Hertog04b} confirms this picture, showing that in the canonical ensemble hairy black holes are unstable against decay into a standard Schwarschild-AdS black hole.}.

Each branch of hairy black holes, associated with excitations about a particular vacuum in the dual field theory, tends to a spherical static soliton solution $\phi_s(r)$ in the limit $R_e \ra 0$. One would expect, therefore, that the no-scalar-hair theorem should hold in all designer gravity theories which do not admit regular spherical solitons. It would be interesting, however, to be able to characterize designer gravity theories where the no-scalar-hair theorem holds purely in terms of properties of the scalar potential. 

The connection between the validity of the no-hair conjecture and the absence of spherical static scalar solitons also means that the PET {\it does} exclude spherical scalar hair for all AdS-invariant designer gravity boundary conditions (\ref{adsW}). This is because if there exists a soliton $\phi_s(r)$, then the conformally rescaled configurations $\phi_\l(r)=\phi_s(\l r)$  provide, for $\l < 1/\sqrt{d-1}$, explicit examples of regular initial data with (arbitrary) negative mass \cite{Hertog06,Heusler92b}. Together with the no-hair results for standard $\a=0$ boundary conditions, this shows that the PET forbids spherical hairy asymptotically AdS black hole solutions in all single scalar field theories where $\phi$ reaches a maximum of $V$ at infinity.

%
\section{Conclusion}
%

We have formulated an extension of the no-scalar-hair theorem for black holes in general relativity, which rules out spherical scalar hair of static asymptotically flat and asymptotically AdS black holes 
if (and only if) the scalar field theory, when coupled to gravity, satisfies the Positive Energy Theorem. This clarifies the status of Wheeler's hypothesis that `black holes have no hair' for e.g. Calabi-Yau compactifications of string theory, where the effective potential typically has negative regions but where supersymmetry ensures the PET holds.
 
The numerical results that we have presented here provide, we believe, strong support for the no-hair conjecture we propose. In particular, we have shown that potentials which are on the verge of violating the PET also separate the set of theories where the no-scalar-hair theorem (and the PET) holds from those where it does not. This applies to all bounded and unbounded scalar potentials with a local minimum at zero, as well as to all potentials with a negative local extremum. When the latter is a local maximum at or slightly above the BF bound, one can adopt a large class of different boundary conditions. Our calculations indicate that the PET excludes spherical scalar hair for all possible choices of AdS-invariant boundary conditions.

If the no-hair conjecture that we propose is proven correct this would mean that the no (spherical) scalar-hair theorem holds in all theories where it can reasonably be expected to hold, and where it is meaningful to even think about it in the first place. Indeed, the significance of the no-hair theorems rests on the validity of the Cosmic Censorship hypothesis, and there is no Cosmic Censorship in theories that do not admit the Positive Energy Theorem \cite{Hertog03b,Hertog03c}. It is also worth noting that the no-hair conjecture we propose here places asymptotically flat spacetimes on equal footing with asymptotically AdS spacetimes\footnote{It appears that the no-hair conjecture should hold even in the presence of a positive cosmological constant. Indeed, asymptotically de Sitter spacetimes do not satisfy the PET and admit hairy black holes \cite{Martinez03}.}. Given the proliferation of hairy AdS black hole solutions in recent years, this comes as an appealing simplification.

We have concentrated on gravity minimally coupled to a single scalar field in four dimensions, but we believe our results - and hence the proposed extension of the no-scalar-hair conjecture - should generalize to scalar multiplets and to nonminimally coupled scalars, possibly in combination with Abelian gauge fields. In the case of a single scalar field it appears that the PET is a sufficient as well as a necessary condition for the no-hair theorem to hold. It would be interesting to see 
whether the conjecture generalizes in both ways to scalar multiplets, where there are potentials satisfying the PET that cannot be written in terms of a superpotential. One would expect there should be a strong connection between the PET and the no-scalar-hair theorem in higher dimensions too, perhaps with some restrictions on the horizon topology. It would also be interesting, of course, to prove the no-scalar-hair conjecture analytically. It seems, however, that none of the methods that have been successfully used to establish no-scalar-hair theorems for positive definite scalar potentials can readily be generalized to potentials with negative regions. 

Finally we have shown that black holes with spherical scalar hair do exist in certain stable designer gravity theories where the boundary conditions on the tachyonic scalar break the asymptotic conformal symmetry. Some of these designer gravity theories have a dual description in terms of a field theory with one or several metastable vacua, and the hairy black holes have a dual interpretation as thermal excitations around one of these vacua. Furthermore, a metastable field theory vacuum itself corresponds to a nontrivial static spherical soliton solution on the gravity side. The fact that in designer gravity one captures the physics of a dual theory with multiple vacua distinguishes these boundary conditions from the usual asymptotically flat or AdS boundary conditions, and provides a qualitative explanation for why the PET does not go together with the no-scalar-hair theorem in designer gravity.

%
\section*{Acknowledgments}
%

I thank Gary Horowitz for discussions and collaboration in the early stages of this project.

\newpage

\appendix

%
\section{Numerical Results}
%

In this Appendix we discuss in more detail the numerical calculations that support the conjecture that 
in theories of gravity coupled to a single scalar with potential $V$, the Positive Energy Theorem holds
if and only if the theory admits no regular static black hole solutions with spherical scalar hair. 
We first consider asymptotically flat black holes, and then turn to potentials with a negative extremum.

\subsection{Asymptotically Flat Black Holes}

First consider potentials of the form 
\be\label{V1app}
V(\phi) = -3 +50 \phi^2 -A \phi^3 + B\phi^6
\ee
where $A$ and $B$ are free parameters. We concentrate on combinations $(A,B)$ for which $V$ has
a local minimum at $\phi_0$ where $V=0$, and a global minimum at $\phi=0$. 
These potentials are represented by the function $B_f(A)$ that separates region II from region III in Figure 2.

\begin{figure}[htb]
\vspace{.2cm}
\begin{picture}(0,0)
\put(46,142){$V$}
\put(205,75){$\phi$}
\put(353,142){$\phi_{e}$}
\put(426,13){$\phi_0$}
\end{picture}
\mbox{
\epsfxsize=7.1cm\epsfysize=5cm \epsffile{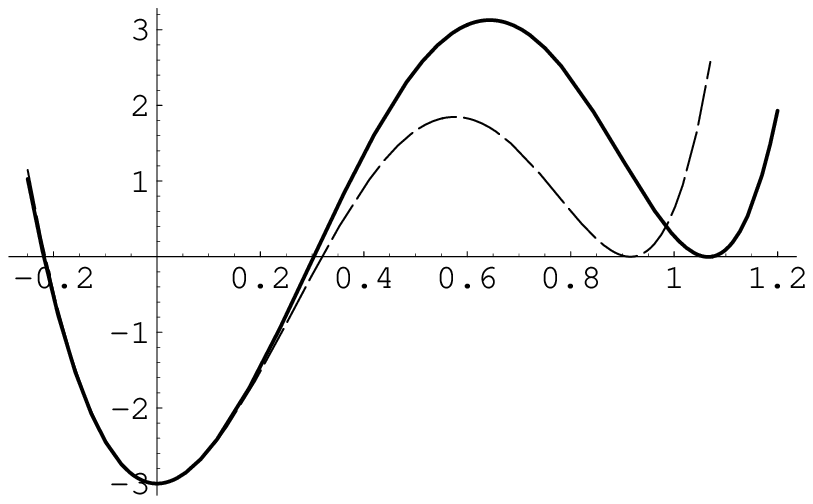} \qquad 
\epsfxsize=6.9cm \epsfysize=5cm \epsffile{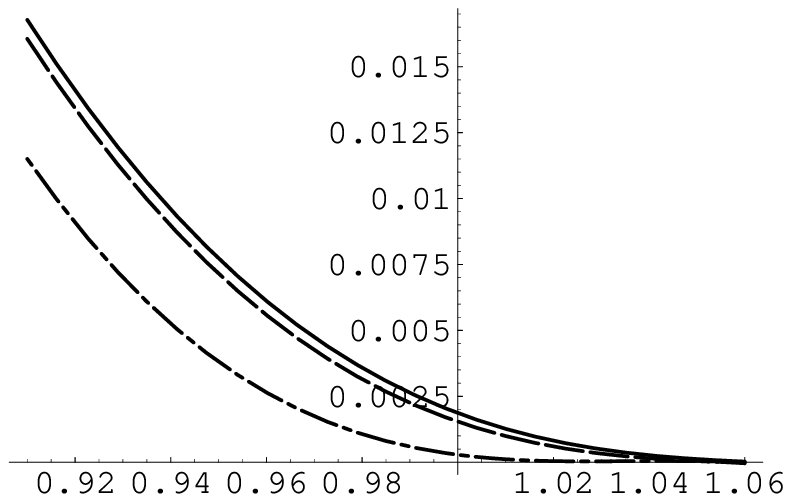}}
\vspace{.2cm}
\caption{The left panel shows a potential (full line) that is on the verge of violating the PET for asymptotically flat solutions. Also shown is a small deformation of this potential (dashed line) that does not satisfy the PET. The right panel shows how $\phi_e$, the value of $\phi$ at the horizon of regular spherical asymptotically flat hairy black holes, changes, for three different values of $R_e$, when one decreases 
$A$ from $A \approx 65$ toward its critical value $A_c=57.6$, while keeping the local minimum at zero. This corresponds to a deformation where the potential changes from the dashed line in the left panel, for which $\phi_0=.92$, to the one given by the full line where $\phi_0=1.06$. The full line in the right panel shows $\phi_e$ as a function of $\phi_0$ for $R_e=5$, the dashed line corresponds to $R_e=2$ black holes and the dot-dashed line to $R_e=1$. One sees that for all radii $R_e$, $\phi_e \ra 0$ when $ A \ra A_c$. There are no regular asymptotically flat hairy black holes for $A <A_c$.}
\label{7}
\end{figure}

For large values of $A$ the theory does not satisfy the PET for configurations where asymptotically
$\phi \ra \phi_0$. The PET holds, however, for $A \leq A_c =57.6$, where $B_f(A)$ joins the 
function $B_c(A)$ of critical 
potentials $V_c$ that are on the verge of violating the PET with asymptotically flat boundary conditions.
In Figure 7 (left panel) we illustrate how the potential deforms when one decreases $A$ from $A \approx 65$
(dashed line) to its critical value (full line). The position of the local extremum at $\phi_0$ increases 
for decreasing $A$ to $\phi_0(A_c) =1.06$ for the critical potential.

We have integrated the field equations (\ref{fieldeq1})-(\ref{fieldeq3}) for static spherical solutions to see 
when theories on $B_f$ admit regular asymptotically flat black hole solutions with scalar hair.
Starting at the horizon for a given radius $R_e$, this amounts to verifying whether there exist a value 
$\phi_e$ so that asymptotically $\phi \rightarrow \phi_0$. One then repeats this procedure for different 
values of $R_e$, and finally for a range of $A$ along $B_f$.
We illustrate the results of this analysis in Figure 7 (right panel), where we plot the value of the scalar
field at the horizon (for which $\phi \rightarrow \phi_0$ at infinity) as a function of $\phi_0$, for several 
radii $R_e$.
One sees that for all radii $R_e$, $\phi_e \rightarrow 0$ precisely when $A \ra A_c$. At the critical point, 
(spherical) hairy black hole solutions cease to exist. Furthermore, we have verified there exists no 
asymptotically flat regular spherical hairy black holes in the regime $A < A_c$ where the PET holds.

\begin{figure}[htb]
\begin{picture}(0,0)
\put(160,144){$V$}
\put(206,75){$\phi$}
\put(262,141){$h$}
\put(435,15){$\phi_e$}
\end{picture}
\mbox{
\epsfxsize=7.1cm\epsfysize=5cm \epsffile{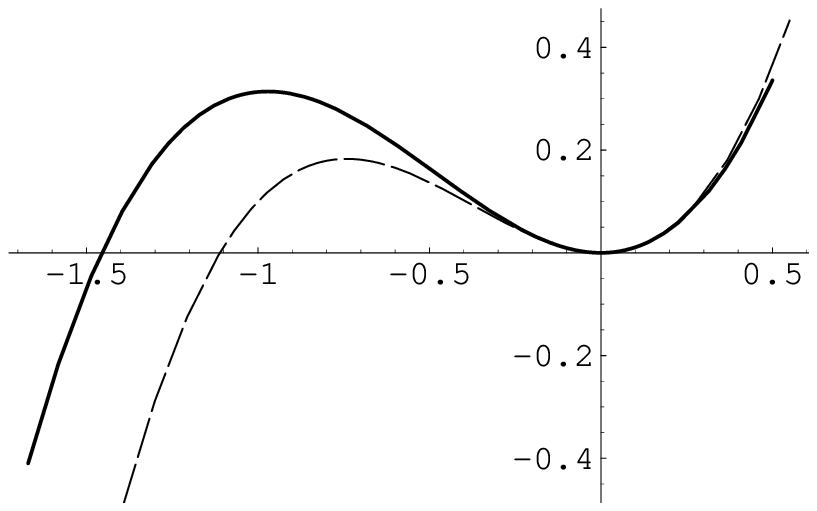} \qquad 
\epsfxsize=6.9cm \epsfysize=5cm \epsffile{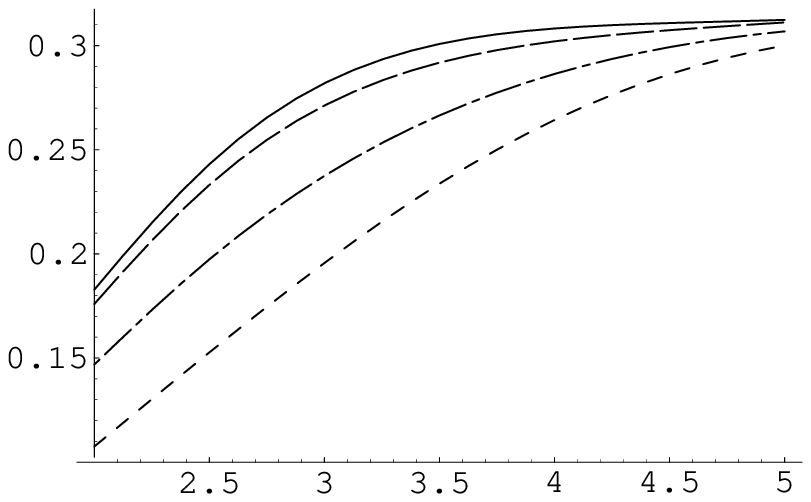}}
\vspace{.5cm}
\caption{The left panel shows a potential (full line) that is unbounded from below, and on the verge of violating the PET for asymptotically flat solutions. The right panel shows how $\phi_e$ changes, for four different values of $R_e$, when one increases the height $h$ of the barrier to its critical value $h_c=.3$. 
The full line in the right panel corresponds to $R_e=10$, the dashed line to $R_e=5$, the dot-dashed line to $R_e=2$ and the dotted line to $R_e=1$. One sees that for all radii $R_e$, $\phi_e \rightarrow \infty$ when the height of the barrier reaches its critical value. For $h > h_c$ there are no regular asymptotically flat hairy black holes.}
\label{8}
\end{figure}

Next we consider a class of potentials that are unbounded from below 
\be \label{V2app}
V(\phi) = \phi^2 + C\phi^3
\ee
Here $C >0$ is a free parameter which controls the height $h=4/27C^2$ of the positive barrier that separates 
the negative region at small $\phi$ from the local minimum at $\phi=0$.
We have seen in section 3 that the PET holds for asymptotically flat boundary conditions if $h > h_c=.3$ 
In Figure 8 (left panel) we illustrate how the potential changes when we increase $h$ towards its critical 
value $h_c$.

Integrating the field equations (\ref{fieldeq1})-(\ref{fieldeq3}) for this set of theories reveals that 
the value of the scalar field at the horizon of regular asymptotically flat hairy black holes becomes
infinitely large when $h$ increases toward its critical valus. We illustrate this in Figure 8 (right panel) 
where we plot the height of the barrier as a function of $\phi_e$, for several radii $R_e$. 
Furthermore, there exist no regular (spherical) hairy black holes for $ h \leq h_c$.

\subsection{Asymptotically AdS Black Holes}

Now we turn to potentials with a negative local extremum, which can be used to specify AdS boundary conditions.

We consider again potentials of the form (\ref{V1app}), but here we concentrate on the subset with
$A=80$. This includes a critical potential $V_c$ for $B_c=43.6$, which has $\phi_0 =.73$ 
and $\Lambda_c = V_c(\phi_0)=-.88$. In Figure 9 (left panel) we illustrate how $V$ changes if we decrease $B$ from $B \approx 46$ (dashed line) 
towards its critical value $B_c$ (full line).

\vspace{.5cm}
\begin{figure}[htb]
\begin{picture}(0,0)
\put(55,142){$V$}
\put(204,70){$\phi$}
\put(310,142){$\phi_e$}
\put(432,14){$\Lambda$}
\end{picture}
\mbox{
\epsfxsize=7.1cm\epsfysize=5cm \epsffile{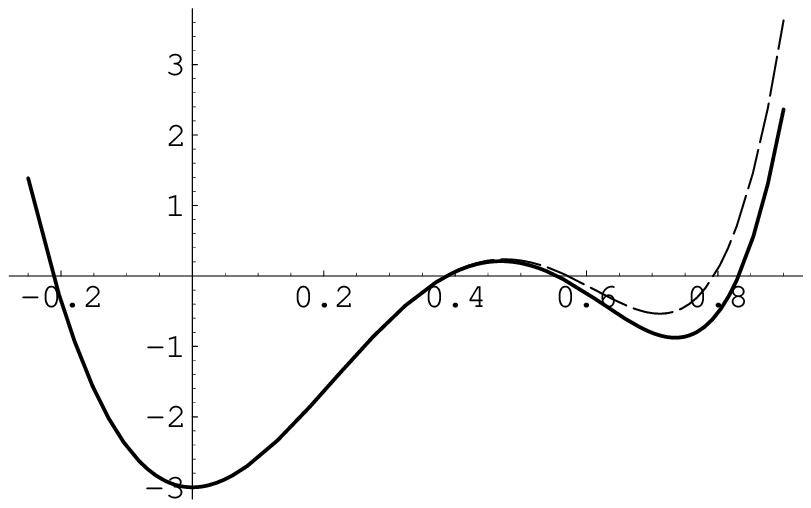} \qquad 
\epsfxsize=6.9cm \epsfysize=5cm \epsffile{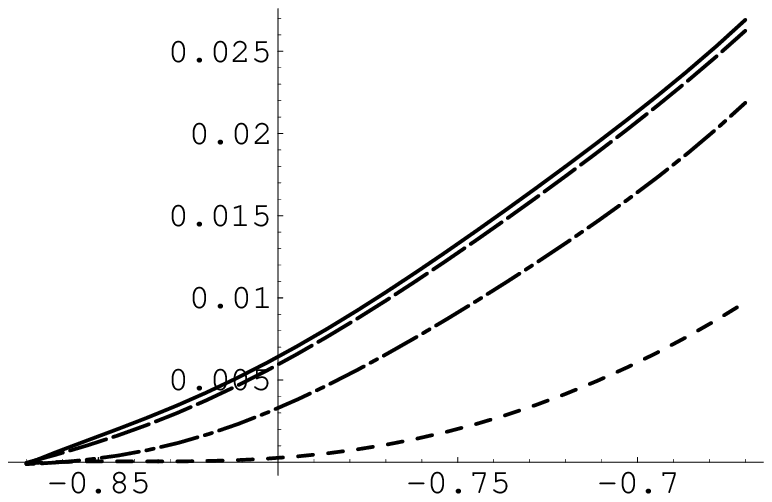}}
\vspace{.5cm}
\caption{The left panel shows a potential (full line) that is on the verge of violating the PET for solutions where $\phi$  asymptotically decays to the local AdS minimum at $\phi_0$. The potential that is given by the dashed line does not satisfy the PET for these asymptotic conditions. The right panel shows $\phi_e$ as a function of $\Lambda = V(\phi_0)$, for four different values of $R_e$, when one decreases $\Lambda$ towards its critical value $\Lambda_c=-.88$. The full line in the right panel corresponds to $R_e=10$, the dashed line to $R_e=5$, the dot-dashed line to $R_e=2$ and the dotted line to $R_e=1$.
One sees that for all radii $R_e$, $\phi_e \rightarrow 0$ precisely when $\Lambda \ra \Lambda_c$. There are no regular spherical hairy black holes where $\phi \ra \phi_0$ at infinity for $ \Lambda < \Lambda_c$ when the false vacuum is stable.}
\label{9}
\end{figure}
\vspace{.5cm}

In Figure 9 (right panel) we show, for several radii $R_e$, the value of the scalar field at the horizon of 
regular spherical hairy black holes as a function of $\Lambda$. As before, one sees that 
$\phi_e \rightarrow 0$ when $\Lambda \rightarrow \Lambda_c$. Furthermore, we find there are no regular 
(spherical) hairy black holes for $B< B_c$ when the AdS solution corresponding to $\phi=\phi_0$ is 
nonperturbatively stable.

To analyse the status of the no-hair theorem for potentials with a negative minimum at $\phi_0$ that are 
unbounded from below, we consider
\be 
V(\phi) = -1+ \phi^2 + C\phi^3
\ee
We have seen in section 3 that even a tiny barrier $ h_c \approx .04$ is sufficient to stabilize the vacuum 
at $\phi=0$. Integrating the field eqs (\ref{fieldeq1})-(\ref{fieldeq3} we again find that $\phi_e \rightarrow \infty$ if 
one raises the height of the barrier to its critical value $h_c$.
We illustrate this in Figure 10 (right panel) where we plot $h$ versus $\phi_e$, for several radii $R_e$.

\vspace{.5cm}
\begin{figure}[htb]
\begin{picture}(0,0)
\put(169,139){$V$}
\put(204,75){$\phi$}
\put(266,139){$h$}
\put(435,85){$\phi_e$}
\end{picture}
\mbox{
\epsfxsize=7.1cm\epsfysize=5cm \epsffile{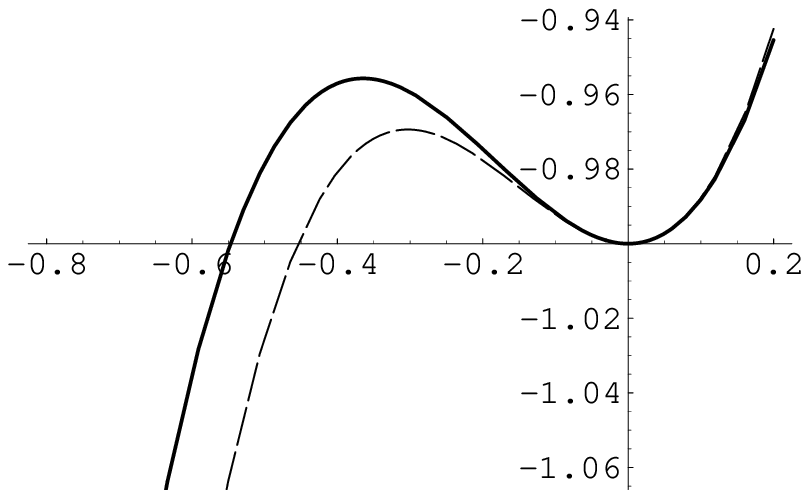} \qquad 
\epsfxsize=6.9cm \epsfysize=5cm \epsffile{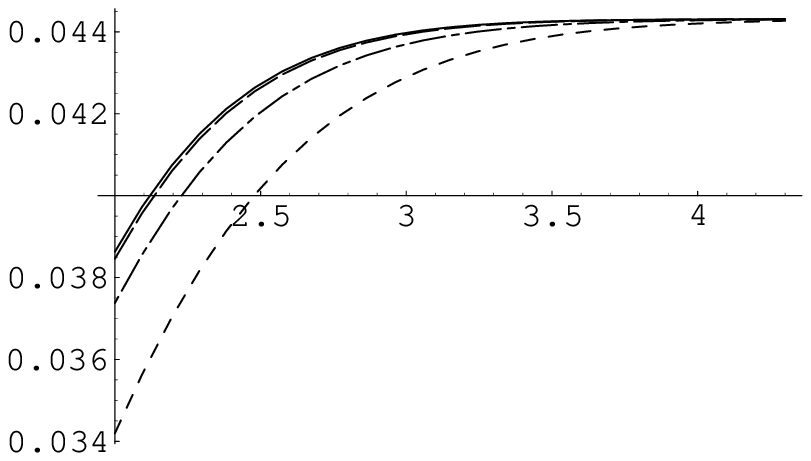}}
\vspace{.5cm}
\caption{The left panel shows a potential (full line) that is unbounded from below, and on the verge of violating the PET for asymptotically AdS solutions. The right panel shows how $\phi_e$ changes, for four different values of $R_e$, when one increases the height $h$ of the barrier to its critical value $h_c=.044$. 
The full line in the right panel corresponds to $R_e=10$, the dashed line to $R_e=5$, the dot-dashed line to $R_e=2$ and the dotted line to $R_e=1$. One sees that for all radii $R_e$, $\phi_e \rightarrow \infty$ when the height of the barrier reaches its critical value. For $h > h_c$ there are no regular asymptotically AdS black holes with spherical scalar hair.}
\label{10}
\end{figure}

\begin{figure}[htb]
\vspace{.5cm}
\begin{picture}(0,0)
\put(109,143){$V$}
\put(205,78){$\phi$}
\put(262,141){$D$}
\put(432,106){$\phi_e$}
\end{picture}
\mbox{
\epsfxsize=7.1cm\epsfysize=5cm \epsffile{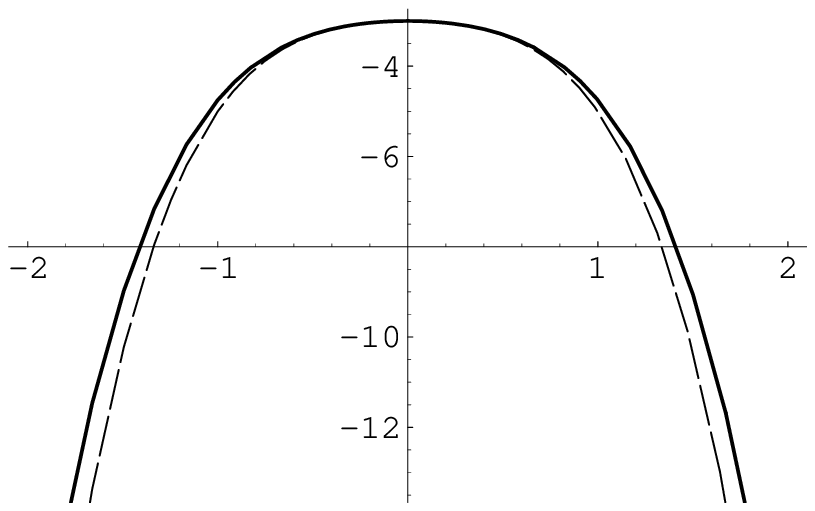} \qquad 
\epsfxsize=6.9cm \epsfysize=5cm \epsffile{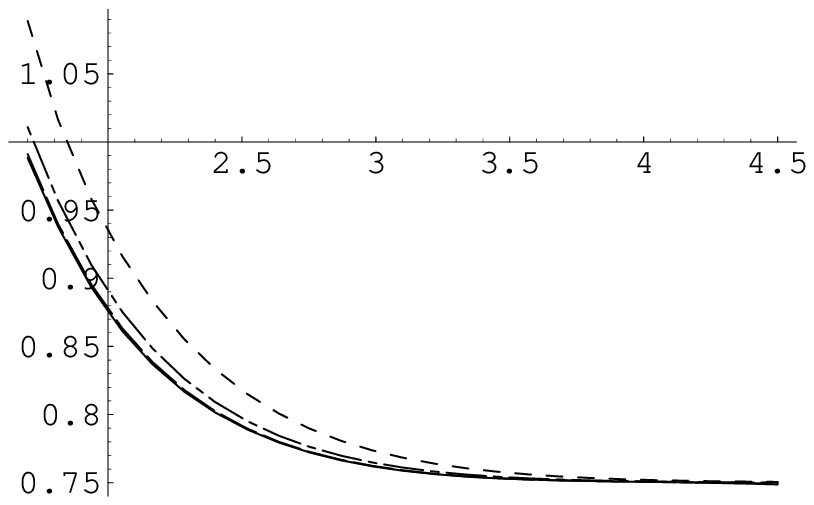}}
\vspace{.5cm}
\caption{The left panel shows a potential (full line) that is on the verge of violating the PET for solutions where $\phi$ asymptotically decays as $\sim 1/r^2$ to the local maximum at $\phi=0$. It has $m^2=-2$ and a critical negative quartic coupling $-D \phi^4$ where $D_c=3/4$. The potential given by the dashed line has $D>D_c$, and hence violates the PET for the same asymptotic conditions. The right panel shows how $\phi_e$, the value of $\phi$ at the horizon of regular hairy black holes where $\phi \sim 1/r^2$ at infinity, changes, for four different values of $R_e$, when one decreases the coupling constant $D$ from $D \approx 1$ to its critical value $D_c$.
The full line in the right panel corresponds to $R_e=10$, the dashed line to $R_e=5$, the dot-dashed line to $R_e=2$ and the dotted line to $R_e=1$. One sees that in all cases, $\phi_e \rightarrow \infty$ when $D \ra D_c$.
For $D <D_c$ the scalar hair of regular asymptotically AdS black holes behaves as 
$\phi \sim 1/r +{\cal O}(1/r^2)$ for all $\phi_e$.}
\label{11}
\end{figure}

We have also discussed theories where $\phi$ approaches a negative maximum of $V$ at 
infinity. Consider the following class of potentials (in $d=4$)
\be
V(\phi) = -3 - \phi^2 - D \phi^4
\ee
where $D$ is a free parameter. The scalar generically decays  as $\phi \sim \alpha/r + \beta/r^2$ at infinity.
If one requires the scalar asymptotically behaves as $\phi \sim r^{-2}$, the PET holds provided $D <3/4$. 
In Figure 11  (right panel) we show that the value of the scalar field at the horizon of regular 
spherical hairy black holes with $\a=0$ becomes infinitely large, for all radii $R_e$, if one 
decreases $D$ towards its critical value. 
Furthermore, for $D<3/4$ the scalar always behaves as a combination of the two asymptotic modes.

\bibliographystyle{JHEP}
\bibliography{ref}


\end{document}